\documentclass[prd,twocolumn,preprintnumbers,superscriptaddress,nofootinbib]{revtex4}
\usepackage[a4paper, hdivide={1.91cm,,1.165cm}, vdivide={1.83cm,,2.6cm}]{geometry}

\usepackage{amstext,amssymb}
\usepackage{amsmath}
\usepackage{graphicx}
\usepackage[hyperfootnotes=false]{hyperref}
\usepackage{xspace}
\usepackage{color}
\usepackage{units}
\usepackage{slashed} 
\usepackage{braket}

\usepackage{wrapfig}
\definecolor{light-gray}{gray}{0.95}
\usepackage{tcolorbox}

\begin{document}

\title{\hspace*{-1.0cm} Study of matter effects in the presence of sterile neutrino using OMSD approximation}

\author{Kiran \surname{Sharma}}
\email{kirans@iitbhilai.ac.in}
\affiliation{Department of Physics, Indian Institute of Technology Bhilai, India}

\author{Sudhanwa \surname{Patra}}
 \email{sudhanwa@iitbhilai.ac.in}
\affiliation{Department of Physics, Indian Institute of Technology Bhilai, India}

\begin{abstract}
We discuss the transition and survival probabilities in $3+1$ neutrino flavor mixing scenario in presence of matter effects. We adopt the well-known OMSD(One Mass Scale Dominance) approximation to carry out our analysis. After that we perform series expansion about $\sin \theta_{13}$ term upto second order. We find that our results are consistent with the already existing $\alpha - \sin \theta_{13}$ approximated relations in the limit of vanishing $\alpha$ and phases involving sterile neutrinos. We also figure out that survival transition probability becomes independent of the fundamental and sterile CP phases under our formalism. Hence, it provides us a new way to look at only matter effects contribution to oscillation probability. Also, the transition probability at the same time gives an independent study of CP-violation arising from the sterile phases, in the vicinity of fundamental CP violation phase. We provide the relation for the   atmospheric probability in the presence of matter by performing the series expansion upto linear order about parameter $A (= 2EV)$, with V being the effective matter potential under OMSD approximation.

\vspace*{-1.0cm}
\end{abstract}

\maketitle

\section{ Introduction}

The existence of neutrino oscillations~\cite{SNO:2001kpb,Super-Kamiokande:1998kpq} emerged as an important tool to look at the beyond standard model physics. A lot of research has already been done to measure the oscillation parameters precisely. The only unknown parameters left to be known accurately in neutrino physics include 
1) The sign of $\Delta m^2_{31}$, depending on if $\Delta m^2_{31} > 0$ the mass hierarchy is considered to be normal hierarchy and for $\Delta m^2_{31} < 0$ it is regarded as an inverted hierarchy. 2) Resolving the octant issue.
3) Determining the amount of leptonic CP violation. In context to the value of $\delta_{CP}$ parameter, the results from T2K~\cite{T2K:2021xwb} and No$\nu$A~\cite{NOvA:2021nfi} experiments show a mismatch. The next generation experiments aim to determine the $\delta_{CP}$ phase at firm footing.

The neutrinos are massless on standard model, but the observation of neutrino oscillations marked neutrinos to be massive. With this the new physics involving the extension of standard model took a revolutionary path. There are several anomalies coming from  the accelerator experiments LSND~\cite{LSND:2001aii} and MiniBooNE~\cite{MiniBooNE:2018esg} and from the so-called reactor~\cite{Mention:2011rk} and Gallium anomalies~\cite{GALLEX:1997lja} that hints towards the existence of a fourth light mass right handed sterile neutrino. The excess of events seen in MiniBooNE has been cross-examined by MicroBooNE~\cite{MicroBooNE:2021ktl,MicroBooNE:2021bcu} which reports no such excess. 

With the emergence of the idea that sterile neutrinos may exist, a wide amount of literature has appeared highlighting the impact of sterile presence on the active neutrino . Various LBL experiments like T2K~\cite{T2K:2019efw}, T2HK~\cite{Agarwalla:2018nlx,Choubey:2017cba}, T2HKK~\cite{Haba:2018klh}, No$\nu$A~\cite{NOvA:2017geg}, DUNE~\cite{Agarwalla:2016xxa} and many more studies~\cite{Gandhi:2015xza,Hollander:2014iha,ANTARES:2018rtf,Kopp:2013vaa} have stressed to provide information on active-sterile oscillations parameters. The new physics provides the subdominant contribution to standard oscillations. The increased number of oscillation parameters as we move from 3 flavor scenario to 3+1 flavor mixing scheme have bring difficulties in understanding the results analytically. The framework developed by authors in ~\cite{Klop:2014ima} have investigated the interference among sterile and atmospheric oscillation frequencies using evolution matrix. We on the other hand, opted an another way to look at the probability expressions in a much simplified way.

We develop a formalism using OMSD which can rule out the simultaneous existence of fundamental CP-phase $\delta_{13}$ and the CP phases arising from sterile neutrinos when working in 3+1 scheme. In this way, this is the first time with the existing literature, where one can study the individual effects of sterile phases analytically. Along with this one can also emphasis the role played by matter effects in determining the exact CP-phase. We perform the series expansion upto linear order in $A$ to compare the transition probability results in vacuum and matter. While doing this analysis, we find an interesting relation for atmospheric probability different from the existing literature.

The paper is structured as follow. In the next section, we provide the theoretical framework. In section 3, we give the analytic formulas for oscillation probabilities using OMSD approximation in sub-section A and then probability expressions are calculated using OMSD and $\sin \theta_{13}$ expansion in sub-section B. We have another sub-section C marking the comparision in vacuum and matter for transition probability $P^{4\nu}_{\mu e}$. Then in section 4, we have our numerical results for 295 km baseline as a case study. The conclusions are mentioned in section 5 followed by an appendix section giving the detail calculations wherever applicable.

\section{Theoretical framework}
\label{sec:probability}
In $3+1$ neutrino oscillation scenario, the flavor eigenstates $\ket{\nu_\alpha}$ (with $ \alpha=e,\mu,\tau,s$) created at $x=0$ with energy $E$ are related to the mass eigenstates $\ket{\nu_j}\,$ (with $j=1,2,3,4$) as,
\begin{eqnarray}
&&\ket{\nu_{\alpha,0}(x=0)} 
\;=\; \ket{\nu_\alpha}
\;=\; \sum_{j=1}^4 U^*_{\alpha j} \ket{\nu_j}\,.
\label{eq:mass-flav}
\end{eqnarray}
where $U$ is the mixing matrix for three active neutrinos and one sterile neutrino which carries information about various mixing angles and phases 
for active as well as sterile neutrinos with following form, 
\begin{eqnarray}
U &=& R\big(\theta_{34}, \delta_{34} \big) \, R\big(\theta_{24}, 0 \big) \, R\big(\theta_{14}, \delta_{14} \big) \nonumber \\
&&\hspace*{2cm} R\big(\theta_{23}, 0 \big) R\big(\theta_{13}, \delta_{13}\big) R\big(\theta_{12}, 0 \big) \nonumber \\
 &\equiv& R\big(\theta_{34}, \delta_{34} \big) \, R\big(\theta_{24}, 0\big) \, R\big(\theta_{14}, \delta_{14} \big) U_{3\nu} 
\end{eqnarray}
where $U_{3\nu}= R\big(\theta_{23}, 0 \big) R\big(\theta_{13}, \delta_{13}\big) R\big(\theta_{12}, 0 \big)$.

At first, the flavor eigenstates are produced and related to mass eigenstates by this transformation. Then after travelling a certain distance the mass eigenstates are transformed to a new flavor eigenstates which can be same or different from the flavor eigenstates produced at source. The evolution of neutrino mass eigenstates can be understood in terms of time dependent Schrodinger equation in the mass basis as,
\begin{eqnarray}
 i \frac{\partial \ket{\nu_j}}{\partial t} &=& H_0 \ket{\nu_j},
  \label{mass_basis}
  \end{eqnarray}
with $H_0$ being the Hamiltonian in the mass basis. It is expressed as a diagonal matrix with energy eigen values as its entries.
   
With the typical value of neutrino energy considered for various neutrino oscillation experiments which is much larger than the neutrino masses, all the neutrinos are considered to be ultra relativistic particles. Under the  extreme relativistic approximations, $t=L$. With the change of basis, the modified form of Schrodinger equation involving flavor eigenstates becomes,
 \begin{eqnarray}
 i \frac{\partial \ket{\nu_\alpha}}{\partial t} \big(t \simeq L \big) &=& \big[H_{4\nu}\big]_{\alpha \beta} \ket{\nu_\beta},
\end{eqnarray}
\noindent
where mass and flavor eigenstates are related by relation given in eq.(\ref{eq:mass-flav}). The effective Hamiltonian in presence of matter can be expressed for $3+1$ neutrino oscillation in flavor basis as,
\begin{eqnarray}
H_{4\nu} \;&=&\; 
\underbrace{
U
\left[ \begin{array}{cccc} 0 & 0 & 0 & 0  \\
                          0 & \Delta m^2_{21}/2E & 0 & 0 \\
                          0 & 0 & \Delta m^2_{31}/2E & 0 \\
                          0 & 0 & 0 & \Delta m^2_{41}/2E  
       \end{array}
\right]
U^\dagger}_{\displaystyle =H_{\rm vac}} \nonumber \\
&+&\underbrace{
\left[ \begin{array}{cccc} V_{CC} & 0 & 0 & 0 \\
                          0 & 0 & 0 & 0 \\
                          0 & 0 & 0 & 0 \\
                          0 & 0 & 0 & -V_{NC}
       \end{array}
\right]}_{\displaystyle =H_{\rm mat}} \;, \nonumber \\
&=& U K U^\dagger + V \, .
\label{H4nu}
\end{eqnarray}
\noindent
This is the master relation accounting the effect of matter in neutrino transition as well as survival probabilities. Here, $K = \mbox{diag}\left(0,k_{21}, k_{31}, k_{41} \right)$ with $k_{i1} = \Delta m^2_{i1}/2E$ is the diagonal matrix carrying information about mass-squared differences and energy. 
One of the important parameters $V_{CC} =\sqrt{2} G_F N_e$ and $V_{NC} =- \frac{1}{2} G_F N_e$ are the effective charged current and neutral current  matter contributions~\cite{Wolfenstein:1977ue,Mikheyev:1985zog,Bethe:1986ej,Smirnov:2004zv} to the total Hamiltonian. Both the potentials are part of effective matter potential matrix relevant for 3+1 scenario. 
The complete diagonalization of effective 3+1 Hamiltonian can be achieved by a new unitary mixing matrix, which will help to derive the oscillations probabilities. Alternatively, we adopt S-matrix formalism, for the derivation of oscillation probabilities. The neutrino flavor propagation can be expressed in terms of evolution matrix as,
\begin{equation}
 \nu_\alpha \big(L \big) = S_{\alpha \beta} \nu_\beta \big(0 \big)\,.
\end{equation}
\noindent
It is to be noted that S-matrix satisfies the same Schrodinger equation and thus, it can be expressed in terms of effective Hamiltonian as,
\begin{eqnarray}
S_{\beta \alpha}
& = &
\braket{\nu_\beta |\nu_{\alpha,0}(x=L)} \phantom{\Bigg|} =
\left[\,
\exp\!\left(-i H_{4 \nu} L\right)
\right]_{\beta\alpha}
\;,
\end{eqnarray}
\noindent
As mentioned in the very beginning, the flavor and mass eigen states are related via unitary mixing matrix.
$\nu_{\alpha} = U_{\alpha i} \nu_{i} $ with $U$ being the standard $4\times4$ parameterized mixing matrix for $3+1$ scenario. The complete formalism of obtaining the S-matrix is achieved in various sub-stages, with the help of changing of basis. At the very first stage, the $4 \times 4$ unitary mixing matrix can be decomposed into two parts, one coming from three active neutrinos only ($U_ {3\nu}$) while the other part is from sterile-active neutrinos ($\overline{U}$) as $U = \overline{U} U_ {3\nu}$. The form of $\overline{U}$ is given by 
$$\overline{U} = R\big(\theta_{34}, \delta_{34} \big) \, R\big(\theta_{24}, 0 \big) \, R\big(\theta_{14}, \delta_{14} \big)$$
Now, we will rotate the basis $\overline{\nu}_\alpha = \overline{U}^\dagger_{\alpha \beta} \nu_\beta$ leading to the decomposition of original Hamiltonian as mentioned in ref~\cite{Klop:2014ima} as,
\begin{eqnarray}
 \overline{H}_{4\nu} &=& \overline{H}^{\rm kin} + \overline{H}^{\rm dyn} \nonumber \\
    &=& U_{3\nu} K U^\dagger_{3\nu} + \overline{U}^\dagger V \overline{U} \,.
\end{eqnarray}
After all these simplifications, the transition probabilities can be expressed in terms of S-matrix evolution operator in the eigenbasis of $\nu_e, \nu_\mu, \nu_\tau, \nu_s$ as
\begin{eqnarray}
\overline{S} \;&=&\; \, \pmb{e}^{- i\, \overline{H}\,L} \approx 
\left[ \begin{array}{cc} \overline{S}_{3\nu} & {\bf 0}_{3 \times 1} \\
                          {\bf 0}_{1 \times 3} &  \exp\!\left(- i\, k_{41}\,L\right)
       \end{array}
\right]
\label{HinMatter}
\end{eqnarray}
where the first term gives the kinematic contribution of neutrino oscillations in vacuum, and the second one explains the nonstandard behaviour in the presence of matter interactions. With the arguments given in ref~\cite{Klop:2014ima}, it can be easily verified that  $k_{41}$ is much bigger than one and much bigger than $k_{21}$ and $k_{31}$ also (4,4) entry of
$\overline{H}_{4\nu}$ is much  bigger than all other elements and at the same time the fourth eigenvalues of $\overline{H}_{4\nu}$ is the largest than other three. As a result, the fourth eigenstate $\overline{\nu}_s$ can be studied independently of others. Hence we can express $4 \times 4$ $\overline{H}_{4\nu}$ as $3 \times 3$ effective Hamiltonian as
\begin{eqnarray}
 \overline{H}_{3\nu} &=& \overline{H}^{\rm kin}_{3\nu} + \overline{H}^{\rm dyn}_{3\nu} 
\end{eqnarray}
which are derived in the new basis of $\big(\overline{\nu_e}, \overline{\nu_\mu}, \overline{\nu_\tau} \big)$. After some simplifications, we get the dynamical contributions as,
\begin{eqnarray}
\overline{H}^{\rm dyn} &=& \overline{U}^\dagger V \overline{U} \nonumber \\
& \simeq &V_C \begin{pmatrix}
1-(1-r) s^2_{14}  & r \tilde{s}_{14} s_{24} & r \tilde{s}_{14} \tilde{s}^*_{34} \\
\dagger  & r s^2_{24}  & r s_{24} \tilde{s}^*_{34}  \\
\dagger  & \dagger  & r s^2_{34}
                           \end{pmatrix}
\end{eqnarray}
For mathematical simplifications we have introduced another parameter $r = -\frac{V_{NC}}{V_{CC}} = \frac{1}{2} \frac{N_n}{N_e}\,$ with typical value $0.5$. It is a nice observation that one can recover the known three flavor effective matter potential matrix by neglecting the contribution containing second order terms in mixing angles involving sterile neutrinos. Additionally, one can use it as an extra contribution to the Hamiltonian ($H_{vac} + H_{SI}$) mimicking NSI effects.

 After that one can get the original flavor basis by using unitary transformation such as
 \begin{eqnarray}
S = \overline{U}\, \overline{S}\, \overline{U}^\dagger 
  = \bigg[ \overline{U} \mbox{exp}\big(- i\, \overline{H}\,L\big) \overline{U}^\dagger \bigg]_{\alpha \beta} 
 \end{eqnarray}
 After $S$-matrix calculation, it is straight forward to calculate the oscillation probability of a particular neutrino flavor $\nu_\alpha$ at source is related to the final neutrino flavor $\nu_\beta$ after traversing a distance $L$ is given by
  \begin{eqnarray}
      &&P^{4\nu}_{\alpha \beta} = P^{4\nu} \left(\nu_\alpha \to \nu_\beta; L \right) 
    = \mid S_{\beta \alpha} \mid^2
  \end{eqnarray}
\noindent 
The main objective of the present work is to carry forward the idea of ref~\cite{Klop:2014ima} but at a much simpler level. The simplification is performed using the one mass scale dominance (OMSD) approximation followed by the expansion upto second order in $\sin \theta_{13}$ in the projected Hamiltonian. The results can be invoked in the complete $4 \times 4$ scenario to get the complete probability relations which is the highlighting thrust of the paper.
\section{Analytic formula for $P^{4\nu}_{\mu e}$ and $P^{4\nu}_{e e}$}
\noindent
We will discuss the appearance and disappearance probabilities in presence of sterile neutrino including matter effects with OMSD approximation~\cite{Choubey:2003yp}. 
The usual way is to expand the effective Hamiltonian in matter in terms of two parameters i) the mass hierarchy parameter $\alpha = \Delta_{21}/\Delta_{31}$ (the ratio of two known mass square differences) and ii) mixing parameter $\sin \theta_{13} = \sin\theta_{13}$ (reactor mixing angle) keeping terms up to second order.

Our analysis differs from the work of~\cite{Akhmedov:2004ny} by treating the projected Hamiltonian in matter using OMSD approximation ($\alpha = 0$) leading to simplified relations, instead of analysis considering $\alpha-\sin \theta_{13}$ approximation. We have consider OMSD approximation to calculate the kinetic part contribution to the total effective projected Hamiltonian. Along with, OMSD approximation, we have simultaneously applied the corrections upto second order in $\sin\theta_{13}$ expansion. In this way, our work differs from the work of~\cite{Klop:2014ima}, where the authors have carried out the analysis using S-matrix formalism with $\alpha$ -$\sin\theta_{13}$ approximation.
\subsection{Probabilities using OMSD approximation}
\noindent
The projected evolution matrix in modified basis is $\overline{S}_{3\nu}$. Now we are at the stage to use the OMSD formalism where we neglect the solar mass square difference $\Delta m^2_{21}$ in comparision with the large mass square difference $\Delta m^2_{31}$ along with vanishing CP-phase $\delta_{13}$ and solar mixing angle $\theta_{12}$. The details of the OMSD formalism can be found in ref~\cite{Choubey:2003yp} and presented briefly in the appendix. The resulting components of $\overline{S}_{3\nu}$ are related to the mass eigenvalues $E_i$ and mixing matrix $\widetilde{U}$ within OMSD approximation as,
\begin{eqnarray}
\big[\overline{S}_{3\nu}\big]_{\alpha \beta} &=& \bigg[ \mbox{exp}\big(- i\, \overline{H}_{3\nu}\,L\big) \bigg]_{\alpha \beta} \nonumber \\
&&\hspace*{-1.5cm}= \bigg[\widetilde{U}_{3\nu}\,\mbox{exp}\big(- i\,\, \text{diag}(E_1, E_2, E_3) \,L\big)  \widetilde{U}_{3\nu}^T
 \bigg]_{\alpha \beta} 
\end{eqnarray}
where, the modified energy eigenvalues in presence of matter are given by
\begin{eqnarray}
&&\hspace*{-1.0cm} E_{1,3} =  \frac{1}{4E}\bigg[\Delta_{31} + A \pm \sqrt{(\Delta_{31} \cos2\theta_{13} - A)^2 + (\Delta_{31} \sin2\theta_{13})^2} \bigg]  \nonumber \\
&&\hspace*{-1.0cm}E_2 = 0\,.
\end{eqnarray}
\noindent
and the new unitary mixing matrix as,
\begin{eqnarray}
 \widetilde{U} = R_{23} R_{13}^M.
\end{eqnarray}
These energy eigenvalues and mixing angles will actually simplify the evolution matrix ($\overline{S}_{3\nu}$) in the projected basis which is the highlighting point of our work. This projected evolution matrix can be invoked in the original $3+1$ evolution matrix $\overline{S}_{4\nu}$. Now we can go back to the original flavor basis under the change of basis $\overline{nu}_\alpha \to nu_\alpha$ and the resulting original evolution matrix ($S$) in terms $\overline{U}$ and $\overline{S}_{4\nu}$ 
is described by the relation,

\begin{equation}
     S = \bar{U} \bar{S} \bar{U}^\dagger \,.
\end{equation}
Using $\bar U_{e2} = \bar U_{e3} = \bar U_{\mu 3} = 0$, the $e\mu$ component of evolution matrix contributing to transition probability $P_{\mu e}^{4\nu}$ is deduced to
\begin{eqnarray}
S_{e\mu} = \bar U_{e1} \left[\bar {U}_{\mu 1}^*  \bar {S}_{ee} +  \bar {U}_{\mu 2}^* \bar {S}_{e\mu}\right] 
+ \bar U_{e4} \bar U_{\mu4}^*\bar S_{ss}\,.
\end{eqnarray}
Since  $\bar S_{ss} =  e^{-i k_{14} L}$ oscillates very fast, 
the associated terms are averaged out by the finite energy resolution of the detector. The detailed derivation has been presented in appendix and the transition probability comes out to be
\begin{eqnarray}
\label{eq:pme_gen}
    P_{\mu e}^{4\nu}  \equiv |S_{e\mu}|^2 &=& |\bar U_{e1}|^2 |\bar U_{\mu 1}|^2  |\bar S_{ee}|^2  \nonumber \\
    \nonumber
    &+&   |\bar U_{e1}|^2 |\bar U_{\mu 2}|^2 |\bar S_{e \mu}|^2\\
    \nonumber
      & +&  2 |\bar U_{e1}|^2  {\rm Re} [ \bar U_{\mu 1}^* \bar U_{\mu 2}  \bar S_{ee} \bar S_{e \mu}^*]\\
      &+&  |\bar U_{e4}|^2 |\bar U_{\mu 4}|^2\,.
\end{eqnarray}
Using OMSD formalism and without having the suppressed contributions involving sterile neutrino mixings in the $\overline{H}^{\rm dyn}$, the resulting analytic expression for transition probability is stated below,
\begin{eqnarray}
P ^{4\nu}_{\mu e} \big({\rm OMSD}\big) &=& \frac{7}{4} \cos^2\theta_{14}\, 
\sin ^2 \theta_{14} \sin ^2 \theta_{24} \nonumber \\
&&\hspace*{-2.5cm}+ \frac{1}{4} \cos^2\theta_{14}\, 
\cos\big(\frac{\Delta^M_{31} L}{2 E}\big)+\frac{1}{2} \cos^2\theta_{14}\, \cos\big(4 \theta ^M _{13}\big)\, 
\sin^2 \big(\frac{\Delta ^M _{31} L}{4E} \big) \nonumber \\
&&\hspace*{-2.5cm}- 2\sin \theta_{14}\, \sin 2\theta_{24}\, \sin\theta_{23}\,
\cos\delta_{14} \sin\big(4 \theta ^M _{13}\big)\,
\sin^2 \big(\frac{\Delta ^M _{31} L}{4E}\big)  \nonumber \\
&&\hspace*{-2.5cm}
- 2\sin \theta_{14}\, \sin 2\theta_{24}\, \sin\theta_{23}\,
\sin\delta_{14} \sin\big(2 \theta ^M _{13}\big) \sin\big(\frac{\Delta ^M _{31} L}{2E}\big) \nonumber \\
&&\hspace*{-2.5cm}+4 \cos^2\theta_{14}\, \sin^2\big(2 \theta ^M _{13}\big)\,\sin^2 \theta_{23}\,
\sin^2\big(\frac{\Delta ^M _{31} L}{4E}\big)
\end{eqnarray}
Here, the difference between matter induced energy eigenvalues (as given in the appendix) i.e, $E_i-E_j = \Delta m^2_{ij}/(2 E)$~\cite{Choubey:2003yp} read as,
$$\Delta ^M _{31} = \sqrt{(\Delta_{31} \cos 2\theta_{13} - A)^2 + (\Delta_{31} \sin 2\theta_{13})^2}\,.$$
The survival probability expression for $\nu_e$ using the same formalism is derived as follows,
\begin{eqnarray}
\label{eq:pee_gen}
    P_{e e}^{4\nu}  \equiv |S_{ee}|^2 &=& |\bar U_{e1}|^2 |\bar S_{ee}|^2
    +  |\bar U_{e4}|^2 \, \nonumber \\
&&\hspace*{-2.5cm}= \sin^2\theta_{14} 
  + \frac{3}{4} \cos^2\theta_{14} + \frac{1}{4} \cos^2\theta_{14} \,
     \cos\big(\frac{ \Delta^M_{31} L}{ 2 E}\big) \nonumber \\
&&\hspace*{-2.5cm}
  +  \frac{1}{2} \cos^2\theta_{14}\, \cos\big(4 \theta ^M _{13}\big)\,
     \sin^2 \big(\frac{\Delta ^M _{31} L}{4E}\big)
\end{eqnarray}
It is an interesting observation that in the limit of vanishing sterile neutrino mixings ($\theta_{14},\theta_{24},\theta_{34}\approx 0$), the transition and the survival probabilities retain their well known form in 3-flavor scenario under OMSD approximation,
\begin{eqnarray}
 P_{\mu e} &=& \sin^2\theta_{23} \sin^22\theta_{13}^M \sin^2\left(\frac{\Delta_{31}^M L}{4E}\right)\, \\
 P_{ee} &=& 1 - \sin^22\theta_{13}^M \sin^2\left(\frac{\Delta_{31}^M L}{4E}\right)\,.
\end{eqnarray}
%
\subsection{Probability using OMSD and $\sin \theta_{13}$ expansion}
\noindent
We carry out an expansion upto second order about small parameter $\sin \theta_{13} \equiv 0.15$~\cite{CHOOZ:2002qts,K2K:2004tri,MINOS:2010rug}which is supposed to be of order $\epsilon$. Indeed, such kind of expansions have already been explored in literature~\cite{Akhmedov:2004ny} along with the expansion of  another parameter $\alpha = \frac{\Delta_{21}}{\Delta_{31}}$. Our work is different from what has already been explored in the other relevant papers, in the way that our analysis expansion has been performed after applying OMSD approximation, thus ruling out the possibility of applying expansion around $\alpha$ and $\sin \theta_{13}$ parameters simultaneously. The advantage of using such technique, is to simplify the probability analysis in presence of sterile neutrino and at the same time considering the contribution from leading order terms upto second order. Since the fundamental CP phase has been considered zero in OMSD approximation, so the final expressions will involves the CP contributions coming from sterile neutrinos only. As a result, our work can be a solution to rule out the contributions of fake CP violation scenario to the leptonic one~\cite{Cabibbo:1977nk}.
\newline
To look at the contribution of each term present in probability expression of $P^{4 \nu} _{\mu e}$, we have plotted them as a function of $\sin 2\theta_{\mu e}(= 2 \sin \theta _{14} \sin \theta _{24})$ after expanding upto second order in $\epsilon$, in the particular case $\sin \theta _{14} \approx \sin \theta _{24}$.
The dashed curves in FIG~\ref{figure:1} express the quantitative contribution in vacuum while the solid ones give contribution of terms in presence of matter. The light blue curve is from first term (i.e. $ |\bar U_{e1}|^2 |\bar U_{\mu 1}|^2  |\bar S_{ee}|^2$) labelled as $T1$, the green curve represented by $T2$ gives contribution from second term ($|\bar U_{e1}|^2 |\bar U_{\mu 2}|^2 |\bar S_{e \mu}|^2$), $T3$ term shown by blue color for third term ( $2 |\bar U_{e1}|^2  {\rm Re} [ \bar U_{\mu 1}^* \bar U_{\mu 2}  \bar S_{ee} \bar S_{e \mu}^*]$)  and the orange one represents the fourth term ($|\bar U_{e4}|^2 |\bar U_{\mu 4}|^2$) and is labelled as $T4$.

 Remarkably, the major contribution to total probability is coming from $T2$ and $T3$ terms only in the region($\approx 0.03-0.07$) allowed by SBL anomalies. Hence, for carrying out the series expansion in terms of $\sin \theta_{13}$ we consider only the T2 and T3 terms. The expression for transition probability after expanding the terms T2 and T3 upto second order in $\sin \theta_{13}$ are as stated below:
 
 \begin{figure}[t]\centering
	\hspace{0.45pt} \includegraphics[width=0.466\textwidth]{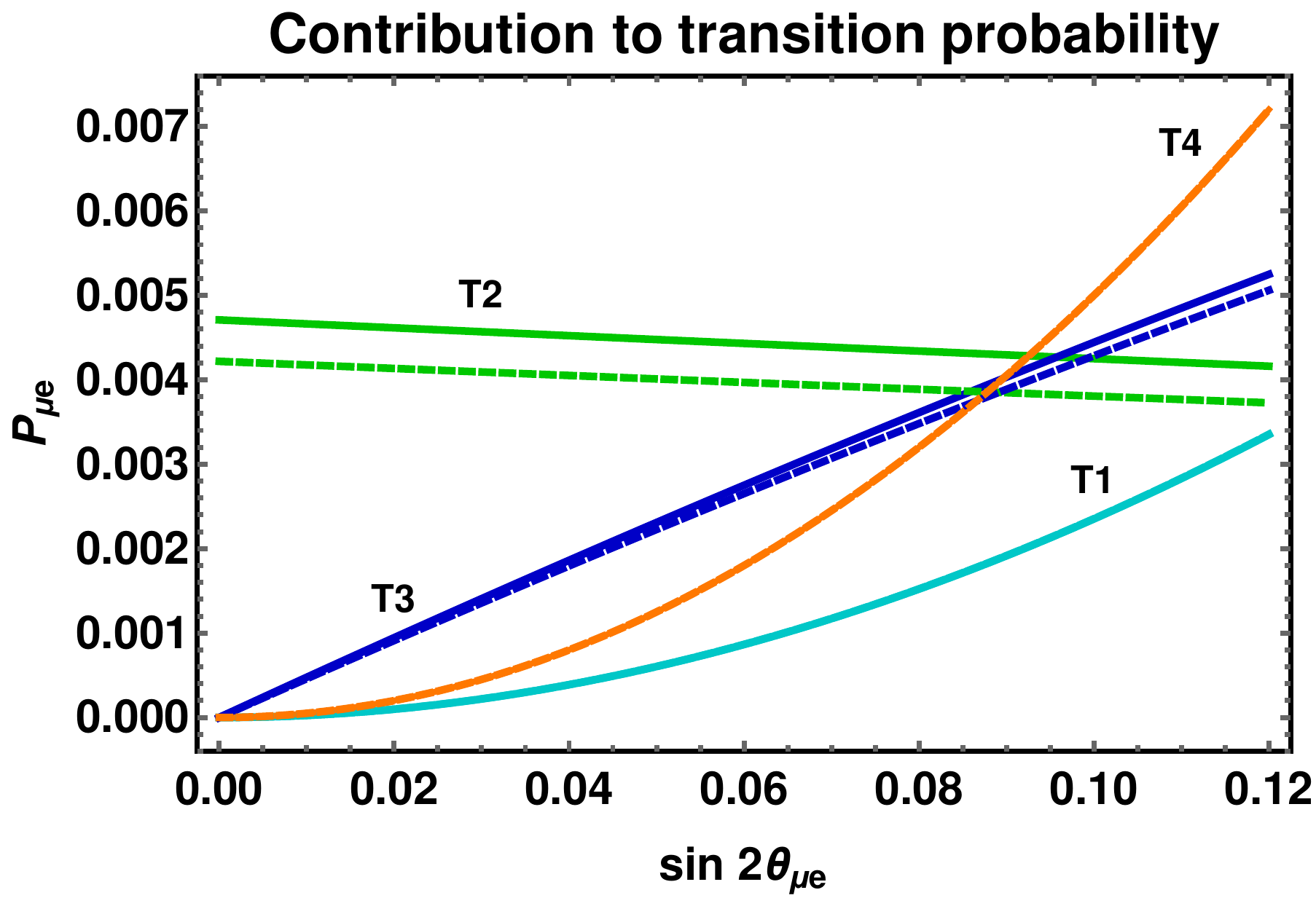}
	\caption{
		The absolute value of four contributions expanded upto $\epsilon ^2$ to the transition probabilities $P_{\mu e}$. The dashed lines represents the behavior of each term in vaccum and the solid one are used for depicting matter effects.} 
	\label{figure:1}
\end{figure}

\begin{eqnarray}
P ^{4\nu (OMSD-s13)}_{\mu e} &=& P ^{4\nu} _{\mu e}(\epsilon ^1) + P ^{4\nu} _{\mu e} (\epsilon ^2)
\end{eqnarray}
\noindent
where the contributions of terms with order $\epsilon ^1$ and $\epsilon ^2$ are as follows
\begin{eqnarray}
P ^{4\nu}_{\mu e}(\epsilon ^1) &=& - \sin\theta_{14} \sin2\theta_{24}\, \sin \theta_{23}\, \nonumber \\
 &&\hspace*{-1.8cm} \times \bigg[2 \cos \delta_{14} \cos ^2 \theta_{14} \times \frac{\sin ^2 \bigg(\frac{\big(\hat{A}-1\big) \Delta _{31}L}{4E}\bigg) }{\hat{A}-1} \nonumber \\
&&\hspace*{-1.8cm} + 2\,\cos ^2 \theta_{14}\, \cos\frac{\big(\hat{A}-1\big) \Delta _{31} L}{4E} \frac{\sin\bigg(\frac{\big(\hat{A}-1\big) \Delta _{31}L}{4E}\bigg) }{\hat{A}-1} \sin \delta_{14} \bigg] \sin \theta_{13} \nonumber \\
P ^{4\nu}_{\mu e}(\epsilon ^2) &=& 4 \cos ^2 \theta_{14} \cos ^2 \theta_{24} \frac{\sin ^2 \bigg(\frac{\big(\hat{A}-1\big) \Delta _{31}L}{4E}\bigg)}{\big(\hat{A}-1\big)^2} \sin ^2 \theta_{23} \sin ^2 \theta_{13} \nonumber 
\end{eqnarray}
\noindent
The survival probability includes the contributions from zeroth order and second order expansion in $\sin \theta_{13}$ i.e.
\begin{eqnarray}
P ^{4\nu (OMSD-s13)}_{e e} &=& P ^{4\nu}_{e e}(\epsilon ^0) + P ^{4\nu}_{e e}(\epsilon ^2)
\end{eqnarray}
where the ${\cal O}(\epsilon^0)$ and ${\cal O}(\epsilon^2)$ dependent terms are given by
\begin{eqnarray}
P ^{4\nu}_{e e}(\epsilon ^0) &=& \frac{3}{4} \cos ^2 \theta_{14} + \frac{1}{4}\cos \frac{(\hat{A}-1) \Delta _{31} L}{4E} \cos ^2 \theta_{14} \nonumber \\
&&+ \frac{1}{2} \cos ^2 \theta_{14} \sin ^2 \frac{(\hat{A}-1) \Delta  _{31} L}{4E} +  \sin ^2 \theta_{14} \nonumber \\
\hspace*{0cm}P ^{4\nu}_{e e}(\epsilon ^2) &=& - \frac{4}{(\hat{A}-1)^2} \cos ^2 \theta_{14} \sin ^2 \frac{(\hat{A}-1) \Delta _{31} L}{4E} \sin ^2 \theta_{13}
\nonumber
\label{survival}
\end{eqnarray}
\noindent
where $\hat{A} = \frac{A}{\Delta_{31}}$ with $A = 2\sqrt{2} G_F N_e E$ as the matter potential contribution. It can be easily verified that in the limit of $\alpha$ tending to zero and in the vanishing limit of the sterile mixing angles, our results are consistent with the series expansion formulas derived in ref~\cite{Akhmedov:2004ny} using $\alpha$ and $\sin \theta_{13}$ expansion upto second order. This is another remark of the present analysis. 

\subsection{\hspace*{-0.5cm} Comparision of probabilities in vaccum and matter}
In the earlier discussion, we find that $T2$ and $T3$ are the dominant contribution to the transition probability as,
\begin{eqnarray}
\label{eq:pme_gen}
P_{\mu e}^{4\nu} &\approx&   |\bar U_{e1}|^2 |\bar U_{\mu 2}|^2 |\overline{S}_{e \mu}|^2
\nonumber \\
      & +&  2 |\bar U_{e1}|^2  {\rm Re} [ \bar U_{\mu 1}^* \bar U_{\mu 2}  \overline{ S}_{ee} \overline{S}_{e \mu}^*] 
\end{eqnarray}
In order to understand how the transition probability for $3+1$ scenario in the presence of matter is related to the vaccum contribution at leading order, we perform a series expansion of $A$ upto linear order. Before that we carry out the series expansion of the modified mixing angle in presence of matter upto linear order in $A$ as,
\begin{eqnarray}
&&\sin^2\,\theta^M_{13} = \sin^2 \theta_{13} \bigg(1 + 2\,A\, \Delta^{-1}_{31} \cos^2 \theta_{13} \bigg) \,, \nonumber \\
&&\cos^2\,\theta^M_{13} = \cos^2 \theta_{13} \bigg(1 - 2\,A\, \Delta^{-1}_{31} \sin^2 \theta_{13} \bigg) \,,
\end{eqnarray}
and the same for the energy eigenvalues,
\begin{eqnarray}
 E_1&=& \frac{\big(1+\cos 2\theta_{13} \big)\,A }{4\,E} \, \nonumber \\
 E_3&=& \frac{\Delta_{31}}{2\,E}\, + \frac{\big(1-\cos 2\theta_{13} \big)\,A }{4\,E} 
\end{eqnarray}
This verifies $E_3-E_1 \simeq \frac{\Delta_{31}}{2\,E} = \big(m^2_3-m^2_1\big)/(2\,E)$. Now we utilise these expression in deriving the projected evolution operator $\overline{S}_{e \mu}$ and $\overline{S}_{e \mu}$. As a result, the simplied leading order (neglecting the $T3$ term) contribution to $P_{\mu e}^{4\nu}$ with the series expansion of $A$ is given by
\begin{eqnarray}
\label{eq:pme_gen}
&&\hspace*{-1.5cm} P_{\mu e}^{4\nu} \approx P_{\rm ATM} \bigg(1+\frac{A}{\Delta_{31}} \cos2\theta_{13} \bigg) \cos^2\theta_{14} \cos^2\theta_{24}\,
\end{eqnarray}
where $P_{\rm ATM} = \sin^2 \theta_{23} \sin\big(2 \theta_{13} \big)\, 
\sin^2 \big(\frac{\Delta_{31}\,L}{4\,E} \big)$ is the known atmospheric dominant contribution. Using $\cos^2\theta_{14} \cos^2\theta_{24} \approx \big(1-2\,\sin^2\theta_{14} -\sin^4 \theta_{14} \big) \simeq \cos^2\theta_{14} \cos^2\theta_{24} \approx \big(1-2\,\sin^2\theta_{14}\big)$, $\cos 2\theta_{13} = 1-2 \sin^2\theta_{13}$ and neglecting the terms containing square or higher order of mixing angles i.e, $\sin^2\theta_{i4}$ or $\sin^2\theta_{13}$, the resulting probability in presence of matter boils down to,
\begin{eqnarray}
\label{eq:pme_gen}
P_{\mu e}^{4\nu} \big({\rm Matter} \big)&& \equiv P^{m}_{\rm ATM}\approx P_{\rm ATM} \bigg(1+\frac{A}{\Delta_{31}}\bigg) 
\end{eqnarray}
The result incorporates the corrections due to matter effects and is different from the known results derived in ref~\cite{Klop:2014ima}. Using the general expression for $A$,
\begin{eqnarray}
A 
& \approx & \left(7.63\times 10^{-5}\,\mathrm{eV}^2\right)
\left(\dfrac{\rho}{\mathrm{g/cm^3}}\right)
\left(\dfrac{E}{\mathrm{GeV}}\right)\;,
\end{eqnarray}
\noindent
and using $\Delta m^2_{31} = 2.55 \times 10^{-3}\,\mathrm{eV}^2$, the averaged matter density $\rho\simeq 2.7\,\mathrm{g/cm^3}$ and the typical first maxima peak energy $E\simeq 0.6$~GeV, we get $\frac{A}{\Delta_{31}} \approx 0.05$ for T2K. 
The difference may be due to the fact that they have used both $\alpha-s13$ approximation~\cite{Akhmedov:2004ny} and perturbation framework~\cite{Arafune:1997hd,Asano:2011nj,Agarwalla:2013tza} to calculate the leading order contribution while we used OMSD approximation followed by series expansion of $A$. The series expansion of sine and cosine functions under our formalism is quite different from the one mentioned in ref~\cite{Klop:2014ima}. Hence, our analytic relation differs from the existing literature. This is another key point of our work.

\begin{table}[ht!]
\begin{center}
\begin{tabular}{lccc}
\hline
\hline
Parameter &  Best Fit values & $1\sigma$   \\
\hline
$\Delta_{21}/10^{-5}~\mathrm{eV}^2 $ (NH ) &7.50  \\
\hline
$\sin^2 \theta_{12}/10^{-1}$ (NH )  & 3.18   \\
\hline
$\Delta_{31}/10^{-3}~\mathrm{eV}^2 $ (NH) & 2.55  \\

\hline
$\sin^2 \theta_{13}/10^{-2}$ (NH) & 2.20  \\

\hline
$\sin^2 \theta_{23}/10^{-1}$ (NH)  & 5.74   \\

\hline
$ \sin^2\theta_{14} $ & 0.02  \\
\hline
$ \sin^2\theta_{24} $  & 0.02  \\
\hline
$ \sin^2\theta_{34} $ &  --   \\
\hline
\hline
\end{tabular}
\caption
{The value of the standard oscillation parameters are taken from the global best fit values quoted in \cite{deSalas:2020pgw}. The value for sterile mixing angles used for carrying out numerical analysis are mentioned alongside.
}
\label{table:1}
\end{center}
\end{table}

\section{Numerical Analysis}

\begin{figure}[t]\centering
	\hspace{0.45pt} \includegraphics[width=0.466\textwidth]{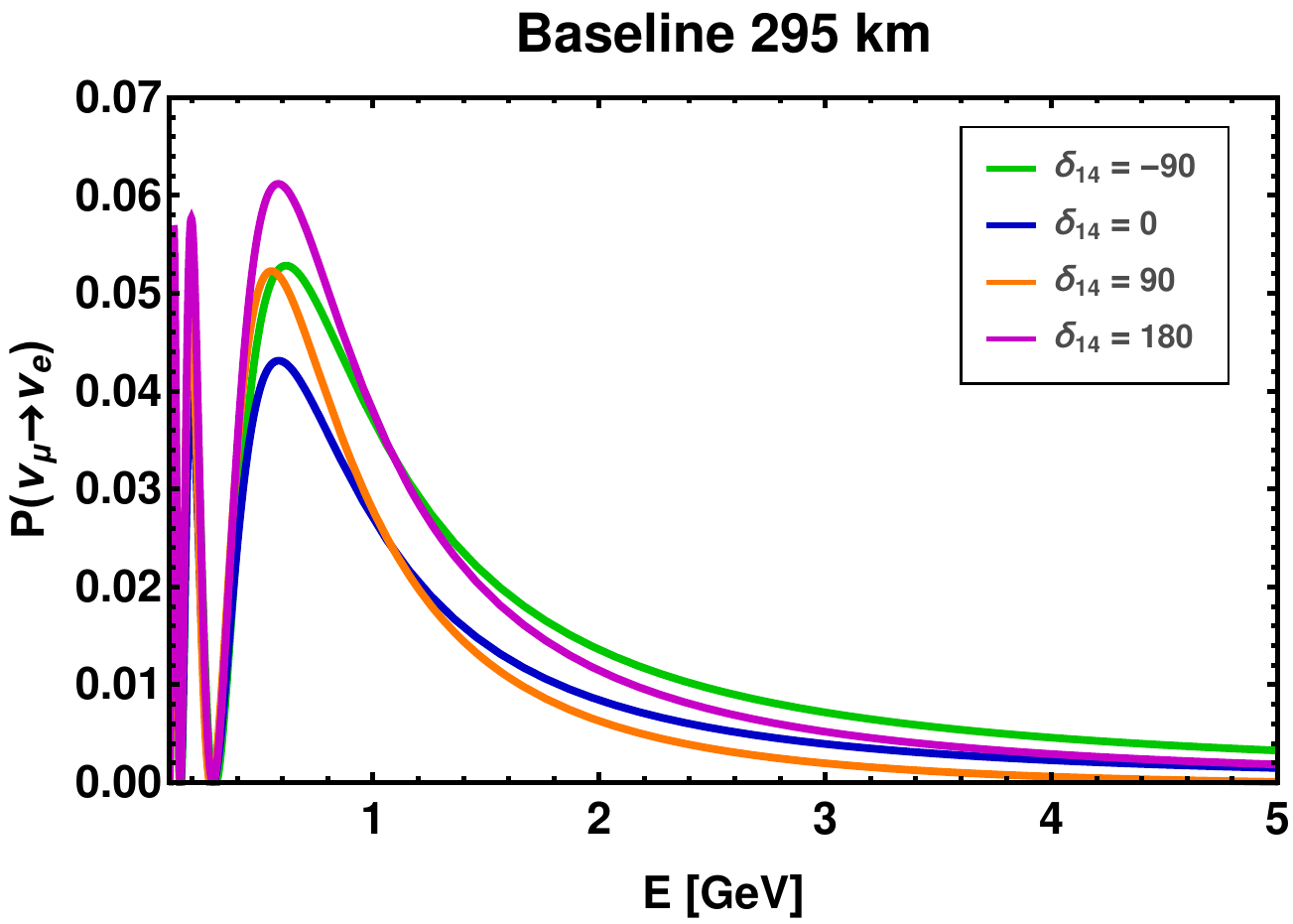}
	\hspace{0.45pt} \includegraphics[width=0.466\textwidth]{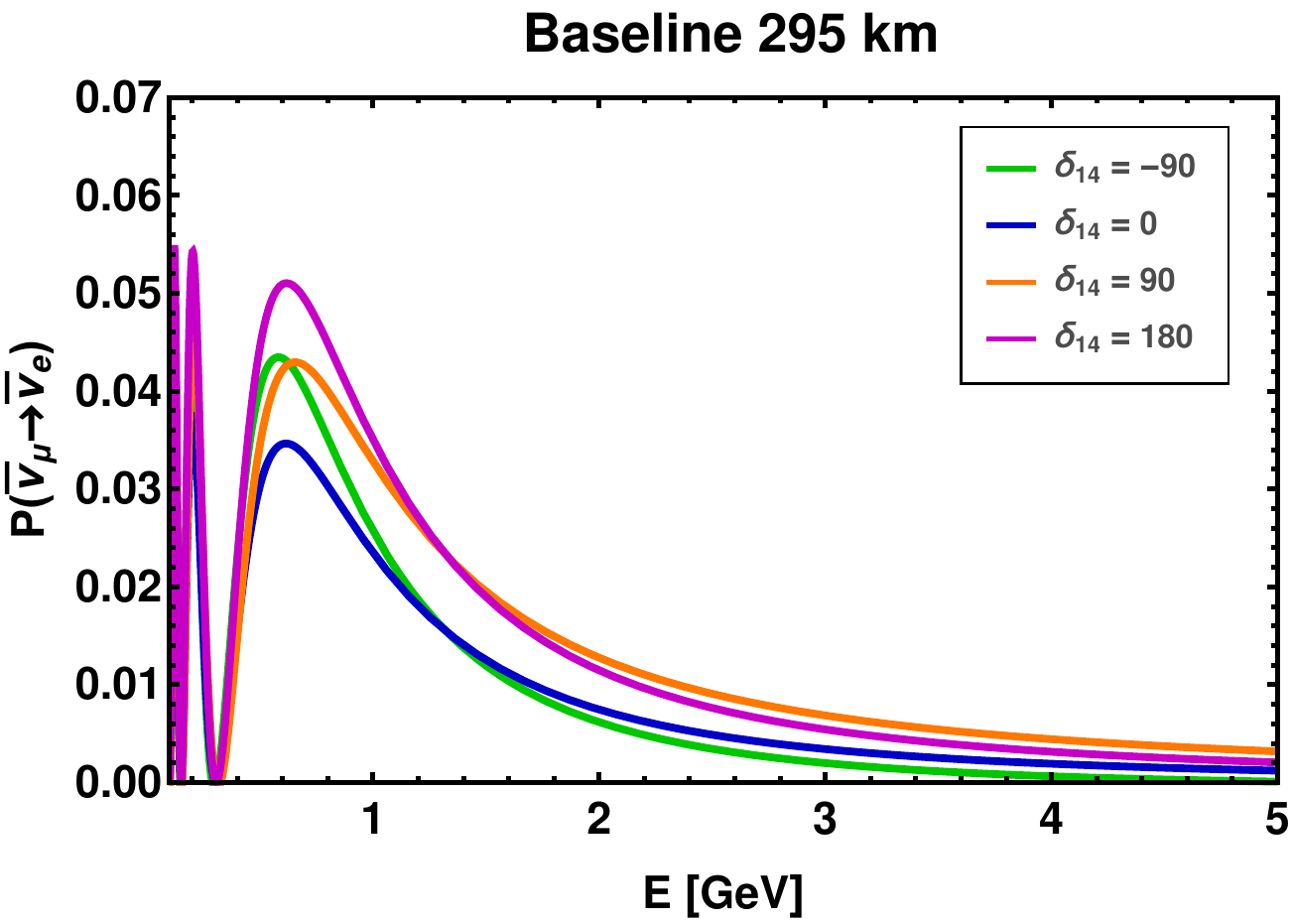}
	\caption{
		The transition probabilities $P_{\nu e}$ for neutrino and anti-neutrino mode as a function of neutrino energy in GeV. The value of $\delta_{14}$ is varied as mentioned in legend. The baseline is kept fixed at 295 km and the matter density $\rho$ is taken as 2.7g/cc.The sterile mixing angles are assigned value $\sin ^2 \theta_{14} = \sin ^2 \theta_{24} = 0.02$} 
	\label{figure:2}
\end{figure}

To look at the validity of the formulas derived under our approach, we perform the numerical analysis for T2K experiment with baseline 295 km. We consider the effect of interaction of neutrinos with matter by assuming a uniform Earth matter density $2.7g/cm ^3$. The hierarchy is considered to be normal through out the numerical calculations. The values of oscillation parameters are taken in reference to the global fit values and are mentioned in table~\ref{table:1} for reference. The value of sterile mixing angles $\theta_{14}$ and $\theta_{24}$ is taken same and is fixed such that $\sin ^2 \theta_{14} = \sin ^2 \theta_{14} = 0.02$. 

In FIG.~\ref{figure:2}, we show the variation of transition probability for neutrinos and anti-neutrinos as a function of energy. The transition probability for antineutrinos is obtained by changing the sign of the MSW potential V and of all the CP-phases, i.e.
\begin{equation}
P_{\overline{\nu}_\alpha \rightarrow \overline{\nu}_\beta} =  P_{{\nu}_\alpha \rightarrow {\nu}_\beta}(\delta_{13} \rightarrow -\delta_{13}, \delta_{14} \rightarrow -\delta_{14}, V \rightarrow -V)
\end{equation}

Since in our case, $\delta_{13}$ is already taken as zero,  the only CP-phase that will change is $\delta_{14}$. The values of CP phase $\delta_{14}$ is varied as shown in legend. The green curve corresponds to $\delta_{14} = -90 ^{\circ}$, the blue curve corresponds to $\delta_{14} = 0 ^{\circ}$, while orange and magenta curves correspond to $\delta_{14} = 90 ^{\circ}$ and $\delta_{14} = 180 ^{\circ}$ respectively. The probability peaks around 0.6 GeV which is the first oscillation maxima for T2K experiment. The mutual swapping of the curves is seen for neutrino and anti-neutrino probability plots as expected.
\begin{figure}[t]\centering
	\hspace{0.45pt} \includegraphics[width=0.466\textwidth]{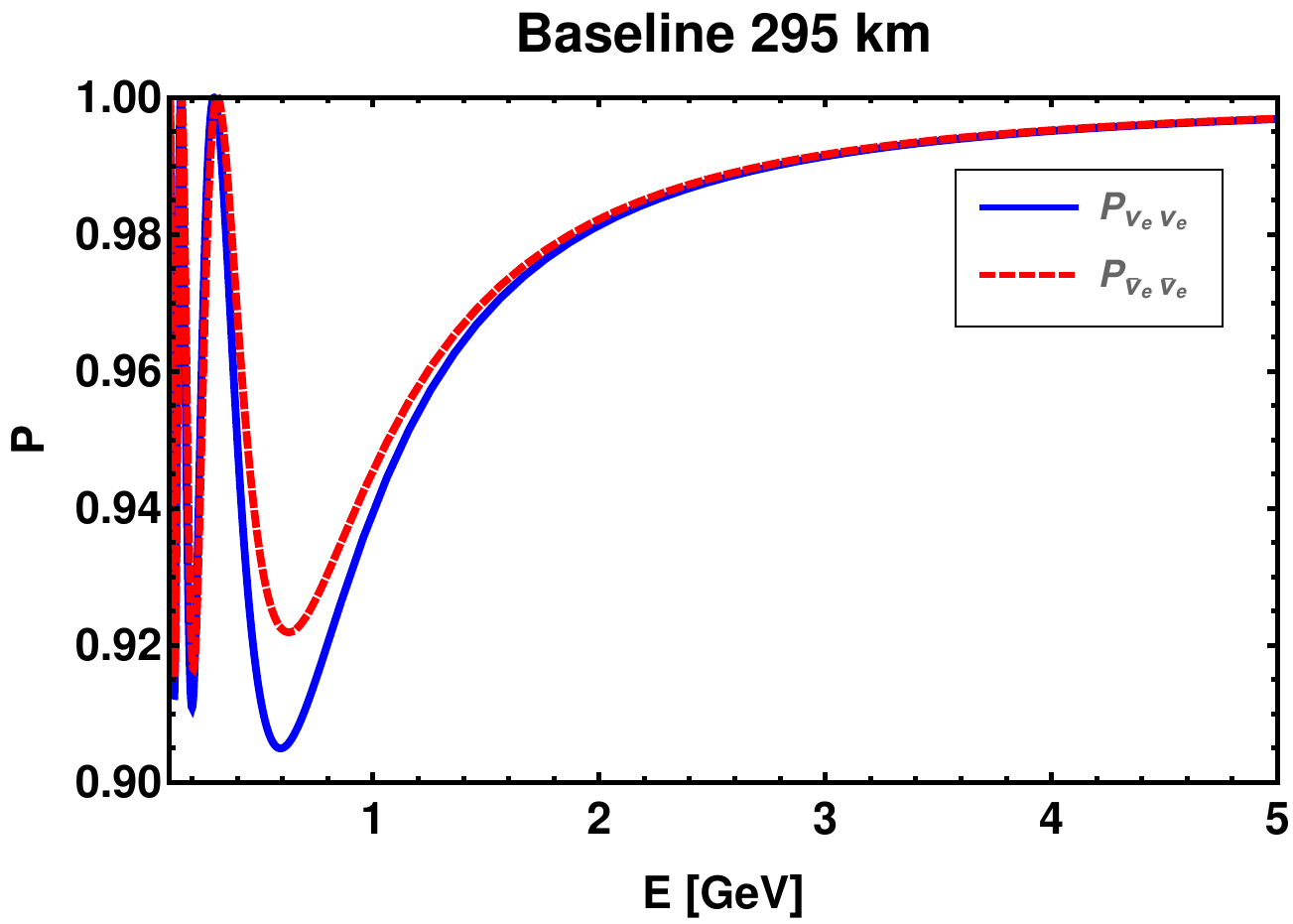}
		\caption{
		The survival probabilities $P_{e e}$ for neutrino and anti-neutrino mode as a function of neutrino energy in GeV. The baseline is kept at 295 km and the matter density $\rho$ is taken as 2.7g/cc. The value of sterile mixing angles is kept constant as $\sin ^2 \theta_{14} = \sin ^2 \theta_{24} = 0.02$ } 
	\label{figure:3}
\end{figure}

\begin{figure}[t]\centering
		\hspace{0.45pt} \includegraphics[width=0.466\textwidth]{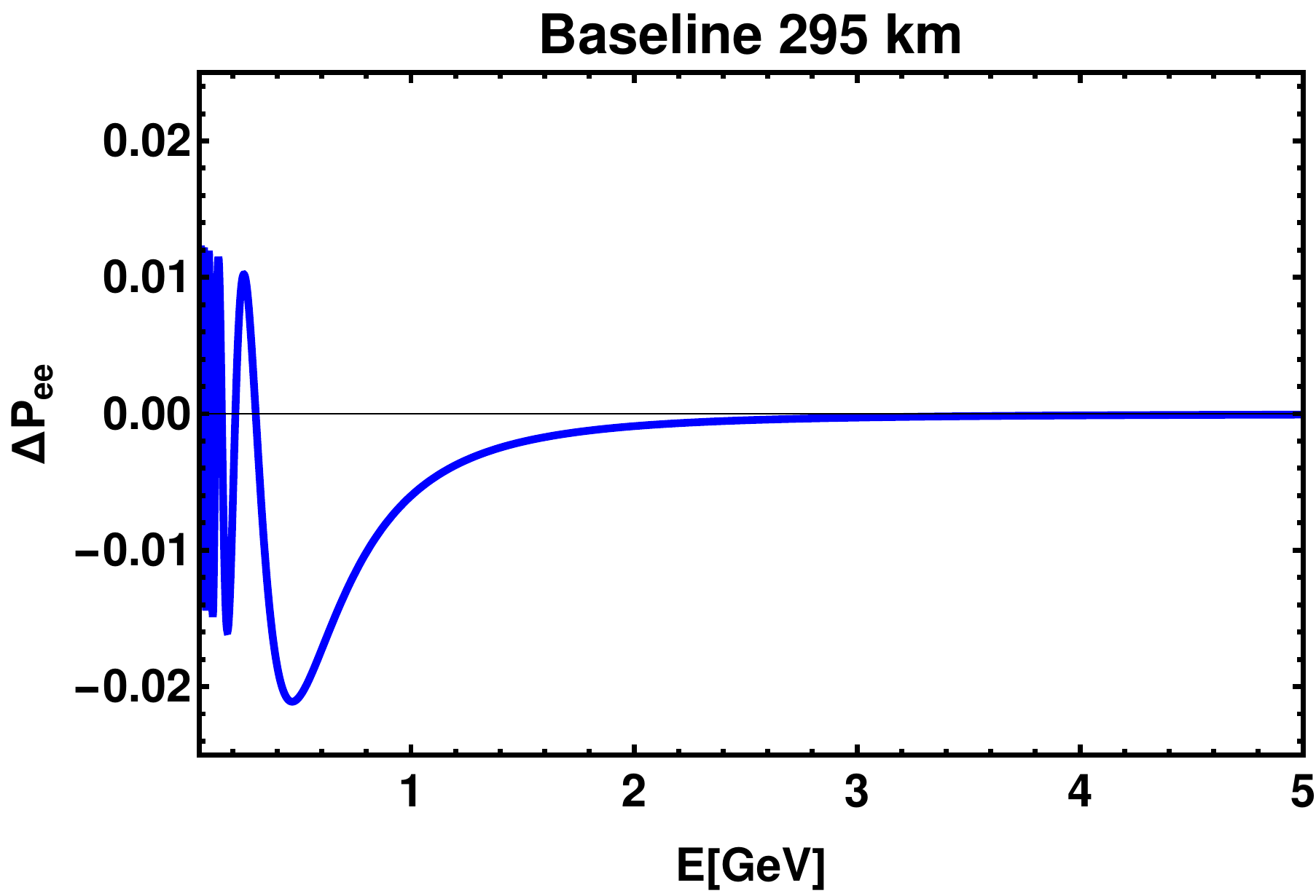}
		\caption{$\Delta P_{ee}$ is shown as a function of energy for 295 km baseline and matter density is kept at 2.7 g/cc.  } 
	\label{figure:5}
\end{figure}

In FIG.~\ref{figure:3}, we look at the variation of survival probability for neutrinos and anti-neutrinos. The solid blue curve represents the probability of neutrino mode and the dashed red curve is drawn by considering anti-neutrinos under study. As evident from eqn \ref{survival}, the disappearance probability $P ({\nu_e} \rightarrow {\nu_e})$ is independent of CP phase $\delta_{14}$. Also as we are using OMSD approximation alongside, the fundamental CP phase i.e.$\delta_{13}$. is also kept zero. Therefore, only a single curve is drawn. A carefull look at the anti-neutrino plot, remarks a little deviation from neutrino probability. This difference $\Delta P_{ee} = P({\nu}_e \rightarrow {\nu}_e) - P(\overline{\nu}_e \rightarrow \overline{\nu}_e)$, between the neutrino and anti-neutrino beams after traversing a finite distance through L in Earth matter has been separately plotted against energy in FIG.~\ref{figure:5}. This finite non-zero difference is arising from the interactions of neutrinos with the matter as they propagate through Earth. Thus, this study shows the significant contribution of matter effects which often leads to create fake CP-violation situation. One thing more interesting to notice is that this non-zero difference between neutrino and anti-neutrino survival probability is maximum around the first maxima and vanishes at higher energy values.

\begin{figure}[b]\centering
	\hspace{0.45pt} \includegraphics[width=0.466\textwidth]{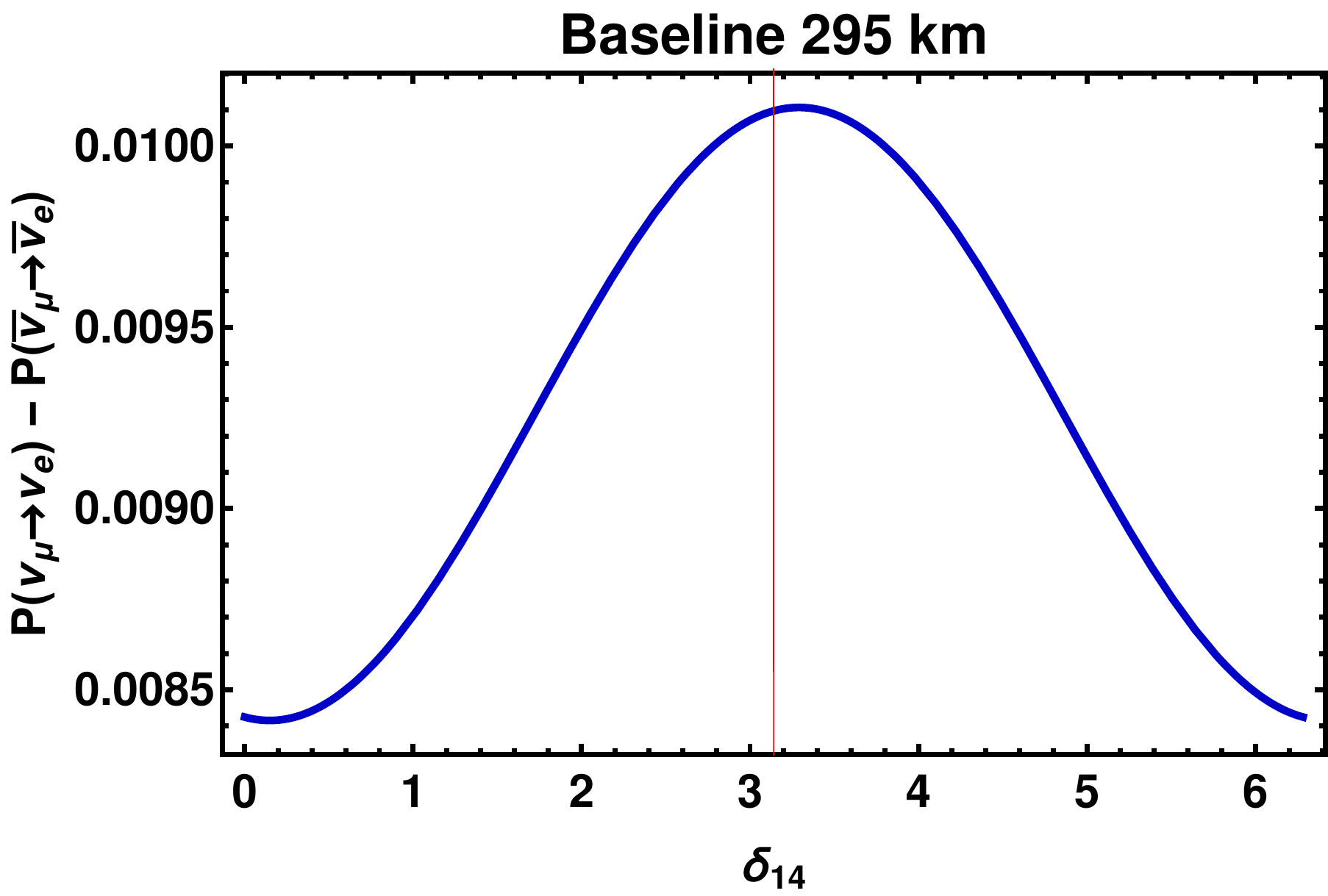}
			\caption{The difference between the neutrino and and anti-neutrino transition probability $P_{\mu \mu}$ as a function of sterile CP-phase $\delta_{14}$. The energy is kept fixed at 0.6 GeV corresponding to the peak value of first maxima for T2K experiment with a baseline of 295 km. The constant matter density $\rho = 2.7 g/cc$ is taken. The value of sterile mixing angles is kept constant as $\sin ^2 \theta_{14} = \sin ^2 \theta_{24} = 0.02$.} 
	\label{figure:4}
\end{figure}

In our analysis of 3+1 scenario the CP violations induced by the new CP-phase $\delta_{14}$ becomes very important. To examine that effect, we look at the variation of difference between survival probability(i.e. $\Delta P_{\mu \mu} = P({\nu}_\mu \rightarrow {\nu}_\mu) - P(\overline{\nu}_\mu \rightarrow \overline{\nu}_\mu)$ ) against CP-phase $\delta_{14}$ in FIG.~\ref{figure:4}.
 We fixed the energy at a constant value 0.6 GeV corresponding to the first oscillation maximum for $P_{\mu e}$. We find that this non-zero difference is strongly related with the sterile CP phase. The red line is drawn for $\delta_{14} = \pi$. A close look at the plot shows that the behavior is not symmetric about this line.

\section{Conclusion}
Our formalism emphasized on bringing in context the relatively simpler analytic formulas for transition and survival probabilities in the $3+1$ scheme. The OMSD method makes our analysis independent of the fundamental CP phase $\delta_{13}$, stressing on the CP-violations induced by sterile phase. The transition probabilities are examined to understand how the different values of $\delta_{14}$ are influencing the oscillation probabilities. The study of survival probability marked an important result as it is independent of sterile CP phase. The non-zero difference between neutrino and anti-neutrino survival probability provides us with another window to look at only matter effects contribution at probability level. Moreover, we have looked at the variation of difference between transition probabilities for neutrinos and anti-neutrinos as they pass through the matter potential as a function of $\delta_{14}$. We provided the relation emphasising the matter corrections to the transition probability involving sterile neutrino in the presence of matter. The numerical analysis have been carried for T2K as a case study and is open to the future LBL experiments to extract more information on new CP violation parameters.

\section{Acknowledgement}
Kiran Sharma would like to acknowledge the Ministry of Education, Govt of India for financial support. KS would like to thank Sabya Sachi Chatterjee for the fruitfull discussions.

\section{APPENDIX}
 
\appendix
\section{Three flavor oscillation probabilities in presence of matter using OMSD approximations:}
The general expression for the oscillation probability~\cite{Giunti:2007ry} for $N$ flavor neutrinos is given by 
\begin{align} \label{N_gen}
 P_{\alpha \beta} = ~ & \delta_{\alpha \beta} - 4 \sum_{i<j} \text{Re} \big( U_{\alpha i} U_{\beta j} U_{\alpha j}^* U_{\beta i}^* \big) \sin^2 \{\Delta_{ij}L/4E\} \\ \nonumber
                  & + 2 \sum_{i<j}  \text{Im} \big( U_{\alpha i} U_{\beta j} U_{\alpha j}^* U_{\beta i}^* \big) \sin\{2 \Delta_{ij}L/4E\},
 \end{align}
 where $\Delta_{ij}=m_i^2 - m_j^2$. 
The standard three neutrino mixing matrix known as Pontecorvo-Maki-Nakagawa-Sakata (PMNS) mixing matrix $U$ and its hermitian conjuagte $U^\dagger$ can be read as,
\begin{eqnarray}
 U &=& \begin{pmatrix}
        U_{e1} & U_{e2} & U_{e3}  \\
        U_{\mu 1} & U_{\mu 2} & U_{\mu 3} \\
        U_{\tau 1} & U_{\tau 2} & U_{\tau 3}  
       \end{pmatrix}  \, , 
  U^\dagger  = \begin{pmatrix}
        U^*_{e1} & U^*_{\mu 1} & U^*_{\tau 1}  \\
        U^*_{e2} & U^*_{\mu 2} & U^*_{\tau 2}  \\
        U^*_{e3} & U^*_{\mu 3} & U^*_{\tau 3}  
       \end{pmatrix} \nonumber
\end{eqnarray}
 
\begin{tcolorbox}[colback=white!10,colframe=yellow!40!black,title=]
\begin{itemize} 
 \item The unitarity condition of the PMNS mixing matrix leads to
 $$\sum^{3}_{j=1} U_{\alpha j} U^*_{\beta j} = \delta_{\alpha \beta}= \bigg\{\begin{array}{cc}
                                                    1 \quad \mbox{if $\alpha = \beta$}\\ 0 \quad \mbox{if $\alpha \neq \beta$}
                                                   \end{array}
  $$
 \item For illustration, consider $\alpha = \mu$ and $\beta=e$, the the off-diagonal unitarity condition becomes
  $$U_{\mu 1} U^*_{e 1} + U_{\mu 2} U^*_{e 2} + U_{\mu 3} U_{e 3} =0\,.$$
  For $\alpha=\mu$, $\beta=\mu$, the relation modifies to
  $$U_{\mu 1} U^*_{\mu 1} + U_{\mu 2} U^*_{\mu 2} + U_{\mu 3} U_{\mu 3} =0\,.$$
\end{itemize}
\end{tcolorbox}
It is difficult to get analytic expressions for three flavor oscillation probabilities which motivates to take various approximation tools to derive simple analytic derivation of appearance and disappearance probabilities. One such approximation is to consider one mass scale dominance (OMSD)~\cite{Cabibbo:1977nk} where the sub-dominant contributions arising from the small solar mass-squared difference i.e, $\Delta m^2_{21} \sim 7 \times 10^{-5} \ \mbox{eV}^2$ is neglected in comparision to the large atmospheric mass-square difference. As a result of this, the effective Hamiltonian for three flavor neutrino oscillation is reduced to,
\begin{eqnarray}
 H_{3\nu} = U 
 \begin{pmatrix}
  0 & 0 & 0 \\
  0 & 0 & 0 \\
  0 & 0 & \Delta_{31}/2E \\
 \end{pmatrix} 
 U^\dagger
 +
\begin{pmatrix}
  \sqrt{2} G_F N_e & 0 & 0 \\
  0 & 0 & 0 \\
  0 & 0 & 0 \\
 \end{pmatrix}.  
\end{eqnarray}
where $\Delta_{31} = \Delta m^2_{31}$, $N_e$ is the number density of the electron background with which neutrino is propagating. It is to be noted that the effects of the solar mixing angle $\theta_{12}$ and of the CP violating phase $\delta_{13}$ in the standard PMNS mixing matrix $U$ become irrelevant and thus, the reduced form of $U$ is given by
\begin{eqnarray}
 U = R_{23} R_{13} = 
 \begin{pmatrix}
1 & 0 & 0 \\
0 & c_{23} & s_{23} \\
0 & -s_{23} & c_{23} \\
 \end{pmatrix}
\begin{pmatrix}
c_{13} & 0 & s_{13} \\
0 & 1 & 0 \\
-s_{13} & 0 & c_{13} \\
 \end{pmatrix}. 
\end{eqnarray}
With this OMSD approximation, one can calculate the modified energy eigenvalues and mixing matrix in presence of matter. The resulting energy eigenvalues of $H_F^{{\rm matt}}$ can be obtained as
\begin{eqnarray}
&&\hspace*{-1.5cm} E_{1,3} =  \frac{1}{4E}\bigg[\Delta_{31} + A  \nonumber \\
    &&\hspace*{-0.3cm} \pm \sqrt{(\Delta_{31} \cos2\theta_{13} - A)^2 + (\Delta_{31} \sin2\theta_{13})^2} \bigg] \\
&&\hspace*{-1.5cm}E_2 = 0.
\end{eqnarray}
Here, the key parameter $A = 2\sqrt{2} G_F N_e E$ is defined for matter contributions to oscillation probabilities. 
The modified mixing matrix due to OMSD approximation in presence of matter becomes,
\begin{eqnarray}
 \widetilde{U} = R_{23} R_{13}^M.
\end{eqnarray}
where the new mixing angle in presence of matter is found to be,
\begin{eqnarray}
 \tan2\theta_{13}^M = \frac{\Delta_{31} \sin2\theta_{13}}{\Delta_{31} \cos2\theta_{13} - A}.
\end{eqnarray}
Using the modified energy eigenvalues and mixing angles, the resulting appearance and disappearance probabilities for three flavor neutrino oscillation using OMSD approximation is given by 
\begin{eqnarray}
 P_{ee} &=& 1 - \sin^22\theta_{13}^M \sin^2\left(\frac{\Delta_{31}^M L}{4E}\right), \\
 P_{\mu e} &=& \sin^2\theta_{23} \sin^22\theta_{13}^M \sin^2\left(\frac{\Delta_{31}^M L}{4E}\right), \\
 P_{\mu \mu} &=& 1 - \cos^2\theta_{13}^M \sin^22\theta_{23} \sin^2\left(\frac{\Delta_{31} + A + \Delta_{31}^M}{4E}\right)L \nonumber \\ \nonumber
             && ~ - \sin^2\theta_{13}^M \sin^2\theta_{23} \sin^2\left(\frac{\Delta_{31} + A - \Delta_{31}^M}{4E}\right)L \\ 
             && ~ - \sin^4\theta_{23} \sin^22\theta_{13}^M \sin^2\left(\frac{\Delta_{31}^M L}{4E}\right),
\end{eqnarray}
with
\begin{eqnarray}
\Delta_{31}^M = \sqrt{(\Delta_{31} \cos2\theta_{13} - A)^2 + (\Delta_{31} \sin2\theta_{13})^2}.
\end{eqnarray}

\section{Oscillation probability for $3+1$ scenario in terms of evolution matrix $S$.}
The formalism of $3+1$ neutrino oscillation can be understood in terms of 
time dependent Schrodinger equation in the mass basis as,
\begin{eqnarray}
 i \frac{\partial \ket{\nu_j}}{\partial t} &=& H_0 \ket{\nu_j},
  \label{mass_basis}
  \end{eqnarray}
with $j=1,2,3,4$ and $H_0$ is defined as the effective Hamiltonian in the mass basis as,
\begin{eqnarray}
H_{0} \;&=&\; 
\left[ \begin{array}{cccc} E_1 & 0 & 0 & 0  \\
                          0 & E_2 & 0 & 0 \\
                          0 & 0 & E_3 & 0 \\
                          0 & 0 & 0 & E_4  
       \end{array} \right]
       \approx E \cdot \left[ \begin{array}{cccc} 1 & 0 & 0 & 0  \\
                          0 & 1 & 0 & 0 \\
                          0 & 0 & 1 & 0 \\
                          0 & 0 & 0 & 1  
       \end{array} \right] \nonumber \\
  &&\hspace*{-0.5cm}     
        + \left[ \begin{array}{cccc} 0 & 0 & 0 & 0  \\
                          0 & \Delta m^2_{21}/2E & 0 & 0 \\
                          0 & 0 & \Delta m^2_{31}/2E & 0 \\
                          0 & 0 & 0 & \Delta m^2_{41}/2E  
       \end{array}
\right]
 \end{eqnarray} 
       and $\nu_j$ being the neutrino mass eigenstate related to flavor eigenstates in following way, 
\begin{eqnarray}
\ket{\nu_j}
\;=\; \sum_{k=1}^3 U_{ j \alpha} \ket{\nu_\alpha}
\end{eqnarray}
With the change of basis, the Schrodinger equation becomes,
 \begin{eqnarray}
 i \frac{\partial \ket{\nu_\alpha}}{\partial t} &=& {[H_{4\nu}]}_{\alpha \beta} \ket{\nu_\beta},
  \label{mass_basis}
  \end{eqnarray}
with $j=1,2,3,4$. In case of $3+1$ scenario, i.e, for three active neutrinos and one sterile neutrino, the mixing matrix $U_{4\nu} \equiv U$ can be parameterized as 
\begin{eqnarray}
U_{4\nu} &=& R\big(\theta_{34}, \delta_{34} \big) \, R\big(\theta_{24}, 0 \big) \, R\big(\theta_{14}, \delta_{14} \big) \nonumber \\
&&\times 
 R\big(\theta_{23}, 0 \big) R\big(\theta_{13}, \delta_{13}\big) R\big(\theta_{12}, 0 \big) \nonumber \\
 &\equiv& R\big(\theta_{34}, \delta_{34} \big) \, R\big(\theta_{24}, 0\big) \, R\big(\theta_{14}, \delta_{14} \big) U_{3\nu}   \nonumber \\
\end{eqnarray}
where $U_{3\nu}= R\big(\theta_{23}, 0 \big) R\big(\theta_{13}, \delta_{13}\big) R\big(\theta_{12}, 0 \big)$ is the standard three flavor neutrino mixing matrix. 

In the present work, we adopt S-matrix formalism~\cite{Akhmedov:2004ny,Asano:2011nj}, in which a particular neutrino flavor changes after traversing a distance $L$ can be defined in terms of an evolution matrix $S$ as
\begin{equation}
 \nu_\alpha \big(L \big) = S_{\alpha \beta} \nu_\beta \big(0 \big)\,.
\end{equation}
The form of evolution matrix can be expressed in term $H_{4\nu}$ as,
\begin{eqnarray}
S_{\beta \alpha}& = &
\left[\,
\exp\!\left(-i H_{4 \nu} L\right)
\right]_{\beta\alpha}
\;,
\end{eqnarray}
The final expression of neutrino oscillation probability $\nu_\alpha$ to $\nu_\beta$ with neutrino energy $E$ and baseline $L$ is expressed in terms of evolution matrix as,
\begin{eqnarray}
P(\nu_\alpha\rightarrow\nu_\beta) \equiv P_{\alpha \beta}
& = &
\bigl|\,S_{\beta\alpha}\,\bigr|^2
\end{eqnarray}

\section{Derivation of $P^{4\nu}_{\mu e}$ and $P^{4\nu}_{\mu e}$ for $3+1$ scenario  using OMSD approximations:}
The simplest way to use OMSD approximation to the full $3+1$ Hamiltonian in presence of matter and solve them to derive the oscillation probabilities. However, this method has limitations of not giving analytic expressions for appearance and disappearance probabilities. Instead, we adopt the analysis carried out by authors in ref ~\cite{Klop:2014ima} where we use the change of neutrino flavor basis wisely such that one can extract the projected Hamiltonian in the basis of three flavor neutrinos and use OMSD approximation in this projected Hamiltonian. Then the projected Hamiltonian and simplified relation for S-matrix using OMSD approximation can be invoked back to $3+1$ scenario for getting analytic results for oscillation probabilities. 

With the change of neutrino flavor basis, we can write down the $3+1$ scenario effective Hamiltonian in presence of matter as,
\begin{eqnarray}
 \overline{H}_{4\nu} &=& \overline{H}^{\rm kin} + \overline{H}^{\rm dyn} \nonumber \\
    &=& U_{3\nu} K U^\dagger_{3\nu} + \overline{U}^\dagger V \overline{U} \,.
    \label{app:Hrnu-bar}
\end{eqnarray}

We would like to apply OMSD approximations in the projected Hamiltonian in the basis of $\overline{\nu}_e, \overline{\nu}_\mu, \overline{\nu}_\mu$. As given in ref.~\cite{Klop:2014ima}, the $3\times 3$ projected Hamiltonian is given as,
\begin{eqnarray}
\overline{H}^{\rm dyn}_{3\nu} & \simeq &V_C \begin{pmatrix}
1-(1-r) s^2_{14}  & r \tilde{s}_{14} s_{24} & r \tilde{s}_{14} \tilde{s}^*_{34} \\
\dagger  & r s^2_{24}  & r s_{24} \tilde{s}^*_{34}  \\
\dagger  & \dagger  & r s^2_{34}
                           \end{pmatrix} \nonumber \\
\nonumber \\
 &&\hspace*{-1 cm} =
 V_C \begin{pmatrix}
1  & 0 & 0 \\
0  & 0  & 0  \\
0  & 0 & 0
                           \end{pmatrix}
                           +
V_C \begin{pmatrix}
-(1-r) s^2_{14}  & r \tilde{s}_{14} s_{24} & r \tilde{s}_{14} \tilde{s}^*_{34} \\
\dagger  & r s^2_{24}  & r s_{24} \tilde{s}^*_{34}  \\
\dagger  & \dagger  & r s^2_{34}
                           \end{pmatrix} \nonumber \\
 &=& \overline{H}^{\rm dyn, LO}_{3\nu}  + \overline{H}^{\rm dyn, NLO}_{3\nu}   
\end{eqnarray}
The $\overline{H}^{\rm dyn, NLO}_{3\nu}$ contains term proportional to $\mathcal{O}\big(\epsilon^2 \big)$ which is contributing to $\overline{S}$. Thus, this contributes to appearance probability as $\mathcal{O}\big(\epsilon^4 \big)$ which can neglected safely. However, these corrections arising from sterile neutrino mixing can be treated equivalent to NSI effects. After these simplifications, the total projected Hamiltonian derived in the basis of $\nu_e, \nu_\mu, \nu_\tau$ is similar to effective Hamiltonian in three flavor neutrino oscillation as,
\begin{eqnarray}
\overline{H}^{\rm }_{3\nu} &=& \overline{H}^{\rm kin }_{3\nu} + \overline{H}^{\rm dyn}_{3\nu}
\end{eqnarray}
The reduced form of projected Hamiltonian in presence of matter using OMSD approximation is already given in eq(A2). 

Using the known mixing elements involving sterile neutrinos, $\overline{U}_{e2} = \overline{U}_{e3} = \overline{U}_{\mu 3} = 0$, the relevant component of evolution matrix $\overline{S}$ is modified as,
 \begin{eqnarray}
  S_{e\mu} &=& \overline{U}_{e1} \bigg[\overline{U}^*_{\mu 1} \overline{S}_{ee}+
    \overline{U}^*_{\mu 2} \overline{S}_{e\mu}\bigg] + \overline{U}_{e4} \overline{U}^*_{\mu 4} \overline{S}_{ss}
        \nonumber \\
    &=& \mathcal{A} + \mathcal{B} \nonumber \\
  S^*_{ e\mu} &=& \overline{U}_{e1} \bigg[\overline{U}^*_{\mu 1} \overline{S}_{ee}+
    \overline{U}^*_{\mu 2} \overline{S}_{e\mu}\bigg]^* + \overline{U}^*_{e4} \overline{U}_{\mu 4} \overline{S}^*_{ss}
        \nonumber \\
    &=& \mathcal{A}^* + \mathcal{B}^*
 \end{eqnarray}

The transition probability is expressed in terms of evolution matrix after these simplifications is as follows,
 \begin{eqnarray}
P^{4\nu}_{\mu e} &=& S_{e \mu} \cdot S^*_{e \mu} \nonumber \\
    &=& \bigg(\mathcal{A} + \mathcal{B} \bigg) \bigg(\mathcal{A}^* + \mathcal{B}^* \bigg)  \nonumber \\
    &=& \mathcal{A} \mathcal{A}^* + \mathcal{A} \mathcal{B}^* 
        + \mathcal{B} \mathcal{A}^* + \mathcal{B} \mathcal{B}^*
 \end{eqnarray}

  Because of fast oscillation due to presence of large mass square difference i.e, $\Delta m^2_{41}$, the absolute magnitude of the term containing $\overline{S}_{ss} = e^{-i\, k_{41} L}$ is averaged out giving a factor of 1/2 as they are proportional to square of the sin or cosine function in the probability expression. In the other hand, the term containing only $\overline{S}_{ss}$ can be averaged out completely from the general expression giving vanishing effect. The terms proportional to $\mathcal{B}$ or $\mathcal{B}$ which involves the fast oscillatory factor $\overline{S}_{ss}$) are vanishing. The non-vanishing term  $\mathcal{B} \mathcal{B}^*$ after averaging out becomes,
 \begin{eqnarray}
  \mathcal{B} \mathcal{B}^* &=&  \big|\overline{U}_{e 4} \big|^2 \big|\overline{U}_{\mu 4} \big|^2 \cdot \frac{1}{2} 
  \simeq \frac{1}{2} \sin^2\theta_{14} \sin^2\theta_{24}\,.
 \end{eqnarray}
 This contribution comes out to be order of $\mathcal{O}(\epsilon^4)$. The left over term contributing to the total probability expression is given by
  \begin{eqnarray}
  \mathcal{A} \mathcal{A}^* &=&  \big|\overline{U}_{e 1} \big|^2 \big|\overline{U}_{\mu 1} \big|^2
     \big|\overline{S}_{ee} \big|^2
      \nonumber \\  
&&+ \big|\overline{U}_{e 1} \big|^2  \overline{U}^*_{\mu 1} 
      \overline{U}_{\mu 2} \overline{S}_{ee} \overline{S}^*_{e\mu}
+
      \big|\overline{U}_{e 1} \big|^2  \overline{U}_{\mu 1} 
      \overline{U}^*_{\mu 2} \overline{S}^*_{ee} \overline{S}_{e\mu}
       \nonumber \\  
   &&+
     \big|\overline{U}_{e 1} \big|^2 \big|\overline{U}_{\mu 2} \big|^2
     \big|\overline{S}_{e\mu} \big|^2
   \end{eqnarray}  
The first term $T_1$ is proportional to $\sin \theta^2_{14}\,\sin \theta^2_{24}$ and thus, its contribution to the total probability is of the order of $\mathcal{O}(\epsilon^4)$. The second and third terms are combiningly contributing to the interference term and can play an important role in determining sterile neutrino parameters. The fourth term is again suppressed by $\mathcal{O}(\epsilon^4)$. 

Thus, the total contributions for transition probability $P^{4\nu}_{\mu e }$ in $3+1$ scenario is found to be,
\begin{eqnarray}
P^{4\nu}_{\mu e} &=& \mathcal{A} \mathcal{A}^* + \mathcal{B} \mathcal{B}^* \nonumber \\
    &=& |\bar U_{e1}|^2 |\bar U_{\mu 1}|^2  |\bar S_{ee}|^2\\
    \nonumber
    &+&   |\bar U_{e1}|^2 |\bar U_{\mu 2}|^2 |\bar S_{e \mu}|^2\\
    \nonumber
      & +&  2 |\bar U_{e1}|^2  {\rm Re} [ \bar U_{\mu 1}^* \bar U_{\mu 2}  \bar S_{ee} \bar S_{e \mu}^*]\\
     \nonumber
      &+&  |\bar U_{e4}|^2 |\bar U_{\mu 4}|^2\, \\
    &=& T1 + T2 + T3 + T4 \,.
\end{eqnarray}
It is to be noted that the contribution form $T1$ and $T4$ are suppressed at least by 
$\mathcal{O}(\epsilon^4)$ and hence, omitted in the total probability calculation.

\bibliographystyle{utcaps_mod}
\bibliography{neutrino.bib}

\end{document}